\documentclass[pre,showpacs,reprint,amsmath,amssymb,superscriptaddress]{revtex4-1}

\usepackage{wasysym}
\usepackage{graphicx}
\usepackage[TABBOTCAP]{subfigure}

\newcommand{\bra}[1]{\ensuremath{\langle{#1}\vert}}
\newcommand{\ket}[1]{\ensuremath{\vert{#1}\rangle}}

\usepackage{color}
\usepackage{soul}

\begin{document}
\title{Critical behavior of hard squares in strong confinement}
 \author{P\'eter Gurin}
 \email[E-mail: ]{gurin@almos.uni-pannon.hu}
 \affiliation{Institute of Physics and Mechatronics, University of Pannonia, P.O. Box 158, Veszpr\'em H-8201, Hungary} 
 \author{Gerardo Odriozola}
 \email[E-mail: ]{godriozo@azc.uam.mx}
 \affiliation{Area de F\'isica de Procesos Irreversibles, Divisi\'on de Ciencias B\'asicas e Ingenier\'ia, Universidad Aut\'onoma Metropolitana-Azcapotzalco,
Av. San Pablo 180, 02200 CD M\'exico, Mexico}
 \author{Szabolcs Varga}
 \email[E-mail: ]{vargasz@almos.uni-pannon.hu}
 \affiliation{Institute of Physics and Mechatronics, University of Pannonia, P.O. Box 158, Veszpr\'em H-8201, Hungary}

\date{\today}

\begin{abstract}
  We examine the phase behavior of a quasi-one-dimensional system of
  hard squares with side-length $\sigma$, where the particles are
  confined between two parallel walls and only nearest neighbor
  interactions occur. As in our previous work (PRE, 94, 050603
  (2016)), the transfer operator method is used, but here we impose a
  restricted orientation and position approximation to yield an
  analytic description of the physical properties. This allows us to
  study the parallel fluid-like to zigzag solid-like structural
  transition, where the compressibility and heat capacity peaks
  sharpen and get higher as $H \rightarrow H_c=2\sqrt{2}-1\approx
  1.8284$ and $p \rightarrow p_c= \infty$. Here $H$ is the width of
  the channel measured in $\sigma$ units and $p$ is the pressure. We
  have found that this structural change becomes critical at the
  $(p_c,H_c)$ point. The obtained critical exponents belong to the
  universality class of the one-dimensional Ising model.  We believe
  this behavior holds for the unrestricted orientational and
  positional case.
\end{abstract}

\maketitle

\section{Introduction}

     The structural and dynamical properties of confined fluids can be
     very different from that of bulk fluids. Upon dimensional
     restriction first order phase transitions transform into
     continuous ones~\cite{%
       Koga_NATURE_2001,%
       Gordillo_JCP_2006,%
       Gordillo_PRE_2009,%
       Stanley_NATURE-PHYSICS_2010,%
       Koga_PNAS_2015%
     }, jamming and glass formation occur~\cite{%
       Bowles_PRL_2009,%
       Bowles_PRL_2012,%
       Bowles_PRL_2013,%
       Godfrey-Moore_PRE_2014,%
       Godfrey-Moore_PRE_2015,%
       Godfrey-Moore_PRE_2016%
     } and significant changes in some transport properties
     arise~\cite{%
       Bechinger_SCIENCE_2000,%
       Kollmann_PRL_2003,%
       Diamant_PRL_2005%
     }. In addition, other phenomena such as wetting, surface
     ordering, and layering transitions may occur in the
     presence of confinement~\cite{%
       vanRoij-Evans_PRE_2001,%
       Enrique_PRL_2005,%
       Enrique_PRE_2006,%
       delasHeras_JCP_2015%
     }. A fundamental issue in confined systems is to understand how
     the dimensional reduction of spatial variables changes the nature
     of phase transitions. It is a well-known result of the van Hove's
     theorem~\cite{%
       vanHove_PHYSYCA_1950,%
       Cuesta_JSP_2004%
     } that particles interacting with short range potentials do not
     exhibit a genuine phase transition in one dimension.  Therefore,
     it is interesting to study how the first order phase transitions
     become continuous in slit-like pores and cylindrical tubes. Along
     this line several molecular fluids and colloidal systems have
     been examined by changing the pore-width in slit-like
     geometries~\cite{%
       Radhakrishnan_PRL_2002,%
       Dijkstra_SOFTM_2014%
     } and the pore-diameter in nanotubes~\cite{%
       Koga_PNAS_2015,%
       Takaiwa_PNAS_2008%
     }. Common results of these studies are that the first order phase
     transitions observed in bulk may become continuous if one of the
     dimensions of the pore is reduced to the order of the size of the
     particle. This happens with the solid-liquid transition of water
     in nanotubes~\cite{%
       Koga_PNAS_2015%
     } and with the isotropic-nematic transition of rod-like colloids
     in slit-like pores~\cite{%
       Dijkstra_JCP_2003%
     }. Confinement may yield new types of phases such as the
     hexatic~\cite{%
       Radhakrishnan_PRL_2002,%
       Engel_PRE_2013%
     } and the biaxial-nematic ones~\cite{vanRoij-Evans_PRE_2001}.
     Moreover, unusual phase transitions can occur such as
     liquid-liquid~\cite{Bianco_SCIREP_2014,Wang_JPCL_2015} and
     smectic-smectic~\cite{Enrique_PRL_2005,Enrique_PRE_2006}.  The
     criticality of confined water has been also studied in slit-like
     pores, where the liquid-liquid phase transition of water
     monolayers terminates at a critical temperature. The observed
     criticality belongs to the universality class of the
     two-dimensional (2d) Ising-model~\cite{Bernard_PRL_2011}.
 
     Along this line, the motivation of our work is to study the
     structural and critical behavior of a 2d system of confined hard
     squares, where the squares are allowed to rotate and move freely
     in a narrow channel of width $H$. In our previous
     work~\cite{Gurin_PRE_2016}, we obtained that a fluid phase with
     parallel alignment to the wall abruptly transforms into a
     solid-like zigzag structure upon increasing pressure.  We have
     found that our system shows extremely similar features than
     others presenting first order phase transitions. These results
     were based on numerical solutions of transfer operator equations
     and simulations, being both in very good accordance with each
     other. It is important to note that with our previous methods the
     sharpness of the structural transition was impossible to study
     for $H \rightarrow H_c$, $H_c$ being the pore width where the
     close packing densities of the parallel fluid and the zigzag
     structures are the same.

     In this paper we further study the above described system in a
     pure analytical way by discretizing the rotational and
     transversal positions. We are employing dimensionless lengths;
     they are understood in $\sigma$ units where $\sigma$ is the side
     length of the square, e.g. the positional coordinates of the
     centers of squares are $x,y=\mathrm{distance}/\sigma$,
     $H=\mathrm{(width\ of\ the\ channel)}/\sigma$ etc. We set $H<2$
     to satisfy the first neighbor interaction condition. This means
     that we have a quasi-one-dimensional system of classical
     particles which can be handled by the transfer operator method.
     It is well known that the \emph{equilibrium} statistical physics
     of our system can be given by solving the following eigenvalue
     equation
\begin{equation}
      \int dy^\prime d\varphi^\prime \,
        K(y,\varphi; y^\prime, \varphi^\prime)\, \psi( y^\prime, \varphi^\prime)
   =  \lambda\, \psi( y, \varphi)
 \; .
\end{equation}
     The kernel $K$ of the above integral operator is given by
\begin{equation} \label{kernel}
      K(y,\varphi; y^\prime, \varphi^\prime)
   =  \frac{e^{- p \sigma_x(y,\varphi; y^\prime, \varphi^\prime)}}{p}
 \, ,
\end{equation}
     where $\sigma_x(y,\varphi; y^\prime, \varphi^\prime)$ is the $x$
     projection of the contact distance of the nearest neighbor
     particles with orientations $\varphi$, $\varphi^\prime$ and
     positions $y$, $y^\prime$; furthermore $\beta=(k_B T)^{-1}$ is
     the inverse temperature and $p$ is the dimensionless longitudinal
     pressure (or force), $p=\beta p_x H \sigma^2$, where $p_x$ is the
     real longitudinal two-dimensional pressure in energy per square
     length units.

     Note that this operator is clearly compact since it fulfills
     $\int dy d\varphi \int dy^\prime d\varphi^\prime\, |K(y,\varphi;
     y^\prime, \varphi^\prime)|^2 < \infty$, i.e. the kernel type is
     Hilbert--Schmidt. Moreover, $K$ (eq.~\eqref{kernel}) is
     everywhere positive and therefore the operator is irreducible.
     Finally, the integral operator is a positive operator in the
     sense that the image of all non-negative functions is non
     negative. For this case, based on the Perron--Frobenius--Jentzsch
     theorem~\cite{vanHove_PHYSYCA_1950,Cuesta_JSP_2004}, it can be
     proved that the dominant eigenvalue ($\lambda_0$) of the operator
     is unique, there is a gap between $\lambda_0$ and the remaining
     part of the spectrum, and $\lambda_0$ is an analytic function of
     $p$ (and $H$)~\cite{Cuesta_JSP_2004}. Thus, phase transitions
     (both, first order and continuous), as traditionally defined in
     the framework of statistical physics, are definitely out of the
     question.

     The aim of the present paper is twofold. On the one hand we want
     to shed some light on the reason why the numerical and simulation
     (or even experimental) studies seem to predict a discontinuous
     behavior in spite of the above cited analytic theory.  On the
     other hand, there still exists the demand to know whether the
     system shows real divergences at least at infinite pressure. To
     this end, we need to define an analytically solvable model, i.e.
     a simpler one, so that the limiting cases can be studied in an
     exact manner. Therefore, here we study a model in which both the
     orientation and the $y$ position of the particles are restricted
     to discretized values and only the $x$ position is continuous. To
     capture the main features of the freely rotating case, we have
     found that the minimal model must have at least four states for
     all particles, see Fig.~\ref{fig:tube}.  Fortunately, this model
     shows qualitatively similar structures to the continuum model.
     Even though the discretization of the orientational and
     positional degrees of freedom rises some fundamental
     questions~\cite{Gurin_PRE_2011}, we believe that the main
     features of the continuous system are captured. We will go back
     to this point at the end of the paper.

     In the next section we present our notation and the analytic
     solution of the discretized model based on the transfer operator
     method. In the third section we examine the behavior of the
     orientational order parameter, correlation lengths and response
     functions, such as the isobaric heat capacity and the isothermal
     compressibility.  Based on these results, we show that the system
     has a 1d Ising-like critical point at infinite pressure and a
     special channel width $H=H_c$. In the light of this result we
     revisit the problem of the structural transition between parallel
     fluid and a zigzag solid-like structures and we conclude that in
     the vicinity of the critical point this transition is practically
     indistinguishable from a genuine first order phase transition.
     Unfortunately, this $H \rightarrow H_c$ study is numerically
     impossible for the freely rotating case~\cite{Gurin_PRE_2016}.
     Finally, in the last section we summarize our results and discuss
     the possible relation with glassy and jammed behavior, mentioned
     frequently in the literature in connection with other
     quasi-one-dimensional hard particle models.

\section{Analytic solution of a simple model}
\label{sec:II}

\begin{figure}[h!]
      \includegraphics{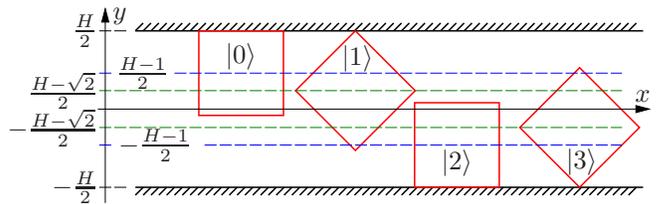}
      \caption{The four possible (allowed) positions of the square
        defined in Eqs.~(\ref{eq:basis}), as labeled.  The particles
        with $\ket{\square}_o$ and $\ket{\Diamond}_o$ orientation can
        move only on the blue and green dashed lines, respectively.
        When a particle changes its orientation, at the same time it
        changes its $y$ position, too.
\label{fig:tube}}
\end{figure}

     In our model the orientation of a particle can have two different
     values: $i)$ the side of the square is parallel to the wall, this
     orientational state of a particle is denoted by $\ket{\square}_o$
     or $ii)$ the diagonal of the square is parallel to the wall,
     which is denoted as $\ket{\Diamond}_o$.  (The index $o$ refers to
     the ``orientation'' and the ``ket'' and ''bra'' symbols follow
     the notation given in Ref.~\cite{Yeomans}.) Furthermore, we
     assume that the particles are always in contact with one wall,
     see Fig.~\ref{fig:tube}.  Thus, if the channel width is $H$ then
     the $y$ coordinate of a particle in the orientational state
     $\ket{\square}_o$ can be $\pm (H-1)/2$, and the $y$ coordinate of
     a particle in the orientational state $\ket{\Diamond}_o$ can be
     $\pm (H-\sqrt{2})/2$.  These $y$ positional states---regardless
     of the difference due to the orientation---are denoted by
     $\ket{+}_y$ and $\ket{-}_y$.

     These four possible combination of orientations and $y$ positions
     form a basis in the four dimensional $\mathcal{H}_y \otimes
     \mathcal{H}_o$ space. These basis states can be written as
\begin{align}
     \ket{0} 
 &:= \ket{+,\square} \equiv \ket{+}_y \otimes \ket{\square}_o
 \nonumber
\\
     \ket{1} 
 &:= \ket{+,\Diamond} \equiv \ket{+}_y \otimes \ket{\Diamond}_o
 \nonumber
\\
     \ket{2}
 &:= \ket{-,\square} \equiv \ket{-}_y \otimes \ket{\square}_o
 \nonumber
\\
     \ket{3}
 &:= \ket{-,\Diamond} \equiv \ket{-}_y \otimes \ket{\Diamond}_o
 \ ,
 \label{eq:basis}
\end{align}
     and the transfer operator acts on the space spanned by the above
     orthonormal basis. The matrix of the transfer operator (the
     transfer matrix) in this basis can be written as \footnote{Here we
       have supposed that $\sqrt{2}/2+1 < H < 2$. The upper bound is
       necessary, but the calculations can be extended without any
       problem in the $\sqrt{2} < H < \sqrt{2}/2+1$ domain when the
       only difference is that $K_{0,3}=K_{1,2}=K_{2,1}=K_{3,0}=e^{-p
         \frac{1+\sqrt{2}}{2}}/p$ which has no any interesting
       consequence. We will see that the $H < 2\sqrt{2}-1$ case is
       unimportant.}
\begin{widetext}
\begin{align}
     K
  =  \frac{1}{p}
     \begin{pmatrix}
        e^{-p} & e^{-p \frac{1+\sqrt{2}}{2}} & e^{-p } &  e^{-p (\frac{3}{2}+\sqrt{2}-H)} \\ 
        e^{-p \frac{1+\sqrt{2}}{2}} & e^{-p \sqrt{2} } & e^{-p (\frac{3}{2}+\sqrt{2}-H)} & e^{-p (2\sqrt{2} - H)} \\ 
        e^{-p} & e^{-p (\frac{3}{2}+\sqrt{2}-H)} & e^{-p } & e^{-p \frac{1+\sqrt{2}}{2}} \\
        e^{-p (\frac{3}{2}+\sqrt{2}-H)} & e^{-p (2\sqrt{2}  - H)} & e^{-p 
\frac{1+\sqrt{2}}{2}} & e^{-p \sqrt{2} }
     \end{pmatrix}.
 \label{eq:K}
\end{align}
\end{widetext}
     Certainly, the operator represented by this matrix is positive,
     irreducible and compact, therefore the theorem presented
     in~\cite{Cuesta_JSP_2004} is valid for this case, i.e.  a phase
     transition is impossible.  The model under study has a simple
     property, namely that there is no ``entanglement'' between the
     orientational and the $y$ degrees of freedom in the sense which
     is clarified below. Therefore, matrix (\ref{eq:K}) can be
     diagonalized in an easy way. For this purpose we define a unitary
     operator $U_y \otimes 1_o$, where $U_y$ acts only on the $y$
     degrees of freedom. Its matrix in the $\ket{+}_y$, $\ket{-}_y$
     basis is
\begin{equation}
      U_y
   =  \frac{1}{\sqrt{2}}
      \begin{pmatrix}
        1 & 1 \\
        1 & -1
      \end{pmatrix},
\end{equation}
     and $1_o$ is the unit operator acting only on the orientational
     degrees of freedom. The unitary operator $U_y \otimes 1_o$ 
uncouples the $\ket{\psi_+}_y := U_y\ket{+}_y = \frac{ \ket{+}_y + 
\ket{-}_y
     }{\sqrt{2}}$ and $\ket{\psi_-}_y := U_y\ket{-}_y = \frac{
       \ket{+}_y - \ket{-}_y }{\sqrt{2}}$ degrees of freedom. This
     means that $K$ has two pieces of two dimensional invariant
     subspaces: $\mathcal{H}_y^+ \otimes \mathcal{H}_o$ and
     $\mathcal{H}_y^- \otimes \mathcal{H}_o$ (where
     $\mathcal{H}_y^\pm$ denotes the one dimensional subspace in
     $\mathcal{H}_y$ generated by the vector $\ket{\psi_\pm}_y$).  In
     other words, the matrix of $K$ after the $U_y \otimes 1_o$
     transformation is a block diagonal matrix, which has the
     following $2 \times 2$ blocks in its diagonal:
\begin{align}
  &\qquad\qquad\qquad\qquad   K_\pm
   =  \begin{pmatrix}
        \Lambda_1^\pm & V^\pm          \\
        V^\pm         & \Lambda_2^\pm
      \end{pmatrix}
   :=
 \label{eq:blocks_K}
\\
  &   \frac{1}{p}\begin{pmatrix}
        e^{-p } \pm e^{-p } & e^{-p \frac{1+\sqrt{2}}{2}} \pm e^{-p (\frac{3}{2}+\sqrt{2}-H)} \\
        e^{-p \frac{1+\sqrt{2}}{2}} \pm e^{-p (\frac{3}{2}+\sqrt{2}-H)} & e^{-p \sqrt{2}} \pm e^{-p (2\sqrt{2}  - H)}
      \end{pmatrix}.
 \nonumber
\end{align}
     The above $K_+$ and $K_-$ symmetric matrices can be diagonalized
     by unitary operators $U_o^{(+)}$ and $U_o^{(-)}$, which act only
     on the orientational degrees of freedom and their matrices (in
     the $\{ \ket{\square},\ket{\Diamond} \}$ basis) can be written as
\begin{equation}
      U_o^{(\pm)}
   =  \begin{pmatrix}
        a_1^\pm & a_2^\pm \\
        a_2^\pm & -a_1^\pm
      \end{pmatrix}
 \label{eq:U_o},
\end{equation}
     where
\begin{equation}
      a_1^\pm
   =  \pm \sqrt{\frac{1 - S^\pm}{2}} 
  \qquad
      a_2^\pm
   =  \sqrt{\frac{1 + S^\pm}{2}}
 \label{eq:a} 
\end{equation}
     and
\begin{equation}
      S^\pm
   =  \frac{\Lambda_2^\pm - \Lambda_1^\pm}{\sqrt{(\Lambda_2^\pm - \Lambda_1^\pm)^2 + (2 V^\pm)^2}}
 \label{eq:S} .
\end{equation}

     Now we can summarize our results detailed above. $K$ can be
     diagonalized by the unitary operator
\begin{align}
     U
 &=  \left( \sum_{s\in\{+,-\}} \ket{\psi_s}_y \bra{\psi_s} \otimes U_o^{(s)} \right) 
     (U_y \otimes 1_o) 
 \nonumber
\\
 &=  \sum_{s\in\{+,-\}} U_y \ket{s}_y \bra{s} \otimes U_o^{(s)},
\end{align}
     that is, the eigenvectors of $K$ are $\ket{\psi_i} = U \ket{i}$
     and the corresponding eigenvalues are $\lambda_i = \bra{i}
     U^\dagger K U \ket{i}$. In detail, the eigenvectors are
\begin{subequations}
\begin{align}
      \ket{\psi_0} 
  &=  \frac{\ket{+}_y + \ket{-}_y}{\sqrt{2}} \otimes 
      \left(a^+_1 \ket{\square}_o + a^+_2 \ket{\Diamond}_o \right)
 \label{eq:psi_0}
 \\
      \ket{\psi_1} 
  &=  \frac{\ket{+}_y + \ket{-}_y}{\sqrt{2}} \otimes 
      \left(a^+_2 \ket{\square}_o - a^+_1 \ket{\Diamond}_o \right)
 \label{eq:psi_1}
 \\
      \ket{\psi_2}
  &=  \frac{\ket{+}_y - \ket{-}_y}{\sqrt{2}} \otimes 
      \left(a^-_1 \ket{\square}_o + a^-_2 \ket{\Diamond}_o \right)
 \label{eq:psi_2}
 \\
      \ket{\psi_3} 
  &=  \frac{\ket{+}_y - \ket{-}_y}{\sqrt{2}} \otimes 
      \left(a^-_2 \ket{\square}_o - a^-_1 \ket{\Diamond}_o \right),
 \label{eq:psi_3}
\end{align}
\label{eq:psi}
\end{subequations}
     and taking into account that $\Lambda_1^- = 0$, the corresponding
     eigenvalues are
\begin{subequations}
\begin{align}
      \lambda_0
  &=  \frac{\Lambda_1^+ + \Lambda_2^+}{2} + 
      \sqrt{  \left( \frac{\Lambda_1^+ - \Lambda_2^+}{2}\right)^2 
            + \left( V^+ \right)^2 }
 \label{eq:lambda_0}
 \\
      \lambda_1
  &=  \frac{\Lambda_1^+ + \Lambda_2^+}{2} - 
      \sqrt{  \left( \frac{\Lambda_1^+ - \Lambda_2^+}{2}\right)^2 
            + \left( V^+ \right)^2 }
 \label{eq:lambda_1}
 \\
      \lambda_2
  &=  \frac{\Lambda_2^-}{2} + 
      \sqrt{  \left( \frac{\Lambda_2^-}{2}\right)^2 
            + \left( V^- \right)^2 }
 \label{eq:lambda_2}
 \\
       \lambda_3
  &=  \frac{\Lambda_2^-}{2} - 
      \sqrt{  \left( \frac{\Lambda_2^-}{2}\right)^2 
            + \left( V^- \right)^2 }.
 \label{eq:lambda_3}
\end{align}
\label{eq:lambda}
\end{subequations}
     It can be seen that for any value of $p$ and $H$ we have $\lambda_0 >
     \lambda_1 > \lambda_2 > 0 > \lambda_3$, however $\lambda_0 >
     |\lambda_3| > \lambda_1 > \lambda_2$.

     As mentioned above, the orientational and $y$ positional degrees
     of freedom are not ``entangled''. We mean that all the
     eigenvectors have a form $\ket{\psi}_y \otimes
     \ket{\psi^\prime}_o$. That is the reason why we need to solve
     only two quadratic equations instead of a quartic one. This
     simple feature of the transfer operator does not hold for the
     freely rotating case. 
     From the general transfer operator theory follows (see
     e.g.~\cite{Yeomans}) that the Gibbs free energy is given by
\begin{equation}
      g
   := \frac{\beta G}{N}
   =  -\log(\lambda_0).
 \label{eq:g}
\end{equation}

     Having a one-particle physical quantity $\mathcal{A}_n$, i.e. for
     a given microscopic state of the system $\mathcal{A}_n$ has four
     (in general) different values, $a(i)$, depending on the state
     only of the $n$-th particle which is labeled by $i$, we can
     define an operator of which matrix is diagonal in the basis given
     by Eqs.~(\ref{eq:basis}): $\bra{i} A \ket{j} = a(i)
     \delta_{i,j}$.  Now the expectation value of $\mathcal{A}_n$
     (which is, in our case, certainly independent of the label of the
     particle, $n$, because we have no positionally dependent external
     fields) can be written as
\begin{equation}
      \langle \mathcal{A}_n \rangle
   =  \bra{\psi_0} A \ket{\psi_0},
\end{equation}
     and the correlation function between $n$-th neighboring particles 
     is given by
\begin{align}
      G_A(n)
 &:=  \langle \mathcal{A}_{m} \mathcal{A}_{m+n} \rangle - \langle \mathcal{A}_{m} \rangle \langle \mathcal{A}_{m+n} \rangle
 \nonumber \\
 & =  \sum_{k \geq 1} \left( \frac{\lambda_k}{\lambda_0} \right)^n
      \bra{\psi_0} A \ket{\psi_k} \bra{\psi_k} A \ket{\psi_0}.
\end{align}

     The orientation and position of a particle give examples for 
     one-particle properties. If we define $\varphi$ as the
     orientation of a particle in state $\ket{\square}$ such as
     $\varphi=\pi/4$ and in state $\ket{\Diamond}$ such as $\varphi =
     0$, then the operator $O$, related to the one particle quantity
     $\mathcal{O}_n = \cos(4\varphi_n)$, can be represented by the
     matrix
\begin{equation}
      \bra{i} O \ket{j} 
   =  \cos(4\varphi_i) \delta_{i,j}
   =  \begin{pmatrix}
        -1  &  0  &  0  &  0  \\
         0  &  1  &  0  &  0  \\
         0  &  0  & -1  &  0  \\
         0  &  0  &  0  &  1  \\
      \end{pmatrix}.
 \label{eq:S_o}
\end{equation}
     Using Eqs.~(\ref{eq:psi_0}), \eqref{eq:a} and \eqref{eq:S} we
     find that the orientational order parameter, $S_o$, which is the
     expectation value of $\mathcal{O}_n$, can be expressed as
\begin{equation}
      S_o
  :=  \langle \mathcal{O}_n \rangle
   =  (a^+_2)^2 - (a^+_1)^2 
   =  S^+,
 \label{eq:<S_o>}
\end{equation}
     and the correlation function as
\begin{equation}
      G_o(n)
   =  (2 a_1^+ a_2^+)^2 \left( \frac{\lambda_1}{\lambda_0} \right)^n,
 \label{eq:G_o}
\end{equation}
     because from Eq.~(\ref{eq:S_o}) it can be seen that $\bra{\psi_2}
     O \ket{\psi_0} = \bra{\psi_3} O \ket{\psi_0} = 0$. In a similar
     way we can define the operator of the $y$ position, denoted also
     by $y$, which can be represented by the matrix
\begin{align}
      \bra{i} y \ket{j} 
  &=  \begin{pmatrix}
         \frac{H-1}{2}  &  0  &  0  &  0  \\
         0  & \frac{H-\sqrt{2}}{2} &  0  &  0  \\
         0  &  0  & -\frac{H-1}{2} &  0  \\
         0  &  0  &  0  & -\frac{H-\sqrt{2}}{2}  \\
      \end{pmatrix}.
 \label{eq:S_y}
\end{align}
     From the above definition we get $\langle y \rangle = 0$ and
\begin{align}
      G_y(n)
  &=  \left(  a_1^+ a_1^- \frac{H-1}{2} 
            + a_2^+ a_2^- \frac{H-\sqrt{2}}{2} \right)^2 
      \left( \frac{\lambda_2}{\lambda_0} \right)^n
    +
 \nonumber\\
  &    \left( a_1^+ a_2^- \frac{H-1}{2} 
            - a_2^+ a_1^- \frac{H-\sqrt{2}}{2} \right)^2 
      \left( \frac{\lambda_3}{\lambda_0} \right)^n.
 \label{eq:G_y}
\end{align}
     Note that the second term alternates sign with $n$ since
     $\lambda_3$ is the only negative eigenvalue.

\section{Analysis of the results}

\begin{figure}[h!]
      \includegraphics{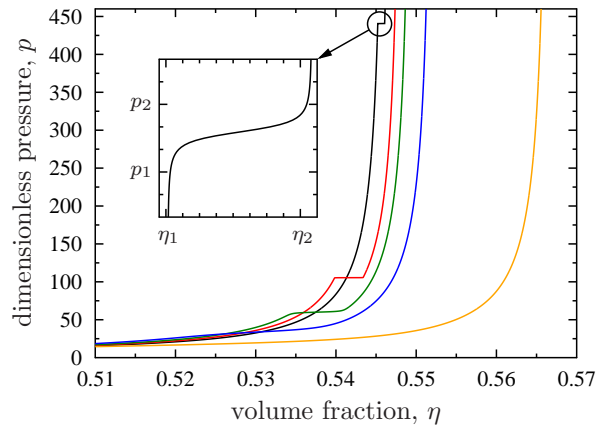}
      \caption{Equations of state for different channel widths.
        Yellow, blue, green, red, and black curves correspond to
        $H=$1.9, 1.85, 1.84, 1.835, and 1.83, respectively. The inset
        zooms in the black curve plateau and highlights its sharpness,
        since $p_2-p_1=3\cdot 10^{-13}$ ($p_1=440.6879559462627$),
        while $\eta_2 - \eta_1 = 8\cdot 10^{-4}$ ($\eta_1=0.5452$).
 \label{fig:eta--p}}
\end{figure}

     Figure~\ref{fig:eta--p} shows the pressure, $p$, as a function of
     the packing fraction, $\eta := N /(H L)$. Here $L$ is the length
     of the channel (normalized by $\sigma$) along the $x$ axis. The
     equation of state can be calculated from the result
     (\ref{eq:lambda_0}) using Eq.~(\ref{eq:g}), yielding
\begin{equation}
      \eta^{-1}
   =  H \frac{\partial g}{\partial p}
 \ .
 \label{eq:eta--p}
\end{equation}

     Although $p(\eta)$ is obtained from an analytic formula, it shows
     a plateau which increases its resemblance with a first order
     discontinuity as $H \rightarrow H_c = 2\sqrt{2}-1 \approx
     1.8284$. Note that for the case of $H=1.83$ (see the inset of
     Fig.~\ref{fig:eta--p}), it already turns almost impossible to
     plot the curve as a smooth function, and this feature worsens for
     $H \rightarrow H_c$.  Therefore, although continuous, the system
     behavior cannot be practically distinguished from a genuine first
     order transition neither by simulations nor---in case this system
     could have an experimental realization---by real experiments.

     The situation is similar to the case of a finite but large ($N
     \approx 10^{23}$) system. We know from the statistical physics
     that in a finite system all derivatives of the free energy are
     continuous. The singularities, in the mathematical sense, appear
     only when considering the thermodynamic limit. However, although
     real systems consist of a finite but very large number of
     particles, experiments clearly show all significant features of
     phase transitions. It is generally accepted that the freezing of
     one liter of water is a genuine phase transition even if the
     system is finite and so there is no mathematical singularity.

\begin{figure}[h!]
   \begin{center}
      \includegraphics{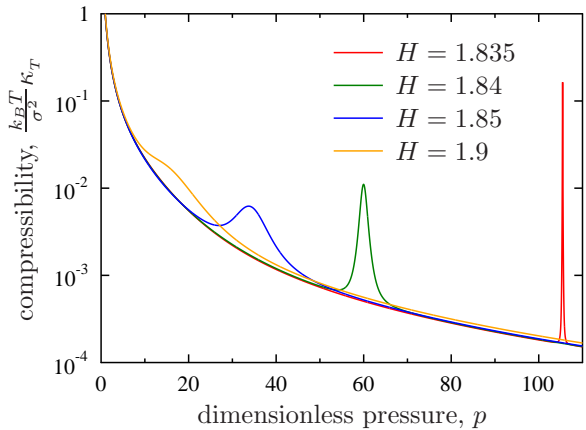}
   \end{center}
   \caption{Dimensionless isothermal (and longitudinal)
     compressibility as a function of (longitudinal and dimensionless)
     pressure.
 \label{fig:compressibility--p}}
\end{figure}

     Before we discuss the reason for our system behavior we show how
     other thermodynamic properties also depict quasi-singularities.
     The isothermal (and longitudinal) compressibility, given by
\begin{equation}
      \kappa_{_T}
  :=  -\frac{1}{L}\frac{\partial L}{\partial p_x}
  \quad \Longrightarrow \quad
      \frac{k_B T}{\sigma^2} \kappa_{_T}
   = -H^2 \eta \frac{\partial^2 g}{\partial p^2},
 \label{eq:kappa}
\end{equation}
     is shown in Fig.~\ref{fig:compressibility--p}.
     As can be seen, $\kappa_{_T}(p)$ peaks at the structural 
     transition, while turning extremely sharp as $H \rightarrow H_c$.

\begin{figure}[h!]
   \subfigure[]
   {  
      \includegraphics{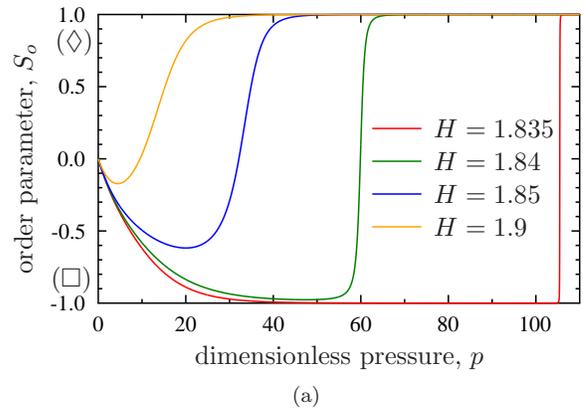}
      \label{fig:<S_o>--p}
   }
   \begin{center}
   \subfigure[]
   {
      \includegraphics{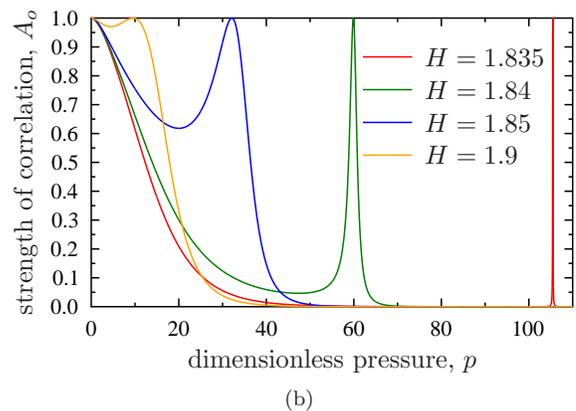}
      \label{fig:A_o--p}
   }
   \subfigure[]
   { 
      \includegraphics{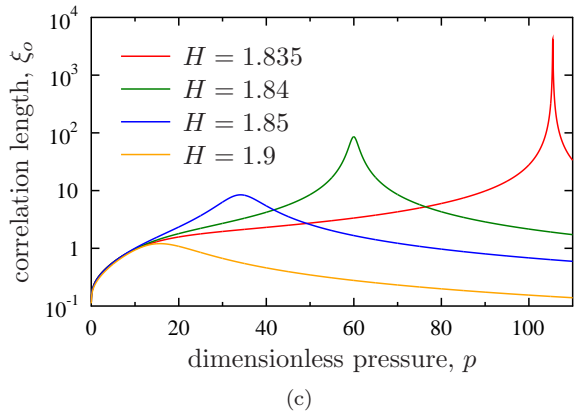}
      \label{fig:xi_o--p}
   }
   \end{center}
   \caption{Orientational order parameter $S_0$ (a),
     strength (b) and length (c) of the
     orientational correlation as a function of the dimensionless
     pressure, $p$. These functions are given by
     Eqs.~(\ref{eq:<S_o>}),~(\ref{eq:A_o}), and~(\ref{eq:xi_o}),
     respectively.
     \label{fig:G_o--p}}
\end{figure}

     Now we examine the order parameters and their correlation
     functions. The orientational order parameter (see
     Eq.~\ref{eq:<S_o>}) can be seen in Fig.~\ref{fig:<S_o>--p}. At
     very low pressure the system behaves as an ideal gas and both
     $\square$ and $\Diamond$ orientations turn equally probable. This
     is simply because we are considering the same number of
     $y$-positions for both $\square$ and $\Diamond$ orientations.
     Thus, this behavior differs from that of the freely
     $y$-positioning system, since in this last case parallel
     configurations are favored by entropy, i.e.  there are more
     configurations for the $\square$-orientation than for the
     $\Diamond$-orientation for an isolated particle. With increasing
     pressure the orientation $\square$ is preferred because the
     $\square$-$\Diamond$ pair at contact has a large $x$-projection
     and the $\Diamond$-$\Diamond$ pair is favored only for large
     $\Diamond$ clusters \cite{Gurin_PRE_2016}.  Then, at a given
     $\tilde p(H)$ pressure a structural change happens in the system
     and the orientation $\Diamond$ becomes more favored. Clearly, the
     $\square$ cannot hold as the packing fraction surpasses
     $\eta_{cp}^{\square}=1/H$, but the structural transition takes
     place slightly below this value.  The orientational correlation
     function, according to Eq.~(\ref{eq:G_o}), can be written as
\begin{equation}
      G_o(n)
   =  A_o e^{-n/\xi_o}
 \label{eq:G_o-exp}
\end{equation}
with 
\begin{equation}
      A_o
   =  (2 a_1^+ a_2^+)^2
 \label{eq:A_o}
\end{equation}
and 
\begin{equation}
      \xi_o
   =  (-\log(\lambda_1/\lambda_0))^{-1}
 \label{eq:xi_o}
\end{equation}

     Note that Eq.~(\ref{eq:G_o-exp}) is not only asymptotically but
     exactly valid for any neighboring distance $n$.  The amplitude of
     the orientational correlations, $A_o$, is depicted in
     Fig.~\ref{fig:A_o--p} and the orientational correlation length,
     $\xi_o$, is shown in Fig.~\ref{fig:xi_o--p}.
     Te orientational correlation length increases monotonically with
     increasing pressure at small densities, but the amplitude of the
     correlation function decreases and the correlations almost vanish
     even for $p \lessapprox \tilde p$ if $H \gtrapprox H_c$. Then,
     the amplitude has a sharp peak when the correlation length is
     maximal, indicating the strong and large scale fluctuations in
     the system at this special point.  Finally, both the amplitude
     and the correlation length go down as $p$ approaches infinity.

     All the above described peculiar behavior happens at a $\tilde
     p(H)$ pressure which can be determined from the $\Lambda^+_1 =
     \Lambda^+_2$ condition. When this equality is fulfilled at high
     pressure, the off-diagonal elements of $K_+$ can be neglected
     (because $V^+ \ll \Lambda^+$ at high $p$). If we suppose, as an
     approximation, that $V^+ = 0$, then the two different eigenvalues
     of $K_+$, namely $\lambda_0$ and $\lambda_1$, cross each other as
     $p$ is increased, and the system yields a first order transition
     at $\tilde p$ from a $\square$ to a $\Diamond$ oriented
     structure. Under this approximation, the Perron--Frobenius
     arguments cannot longer be applied, because the transfer matrix
     has zero entries. Thus, the phase transition appears only by
     forcing the off-diagonal terms in $K_+$ to zero.  By removing
     this imposition, the level crossing of the two eigenvalues is
     avoided, i. e. the two eigenvalues only approach each other and
     the system does not yield a first order transition.  However, the
     closer the eigenvalues approach each other, the more the behavior
     of the system gets reminiscent to that of a phase transition. The
     strength of this reminiscence increases with decreasing $H$ while
     $H>H_c$.  This happens due to the fact that when $H$ approaches
     $H_c$, $\tilde p(H)$ increases and the level crossing
     approximation (the negligence of $V^+$) becomes more valid.
 
     The $\Lambda^+_1 = \Lambda^+_2$ equation can be written, in more
     detail, as
\begin{equation}
      - p (H - H_c) + \log 2 - \log(1 + e^{- p(H - \sqrt{2})})
   =  0
 \label{eq:level_crossing}
\end{equation}
     In case of large pressure ($e^{- p(H - \sqrt{2})} \ll 1$) 
     equation (\ref{eq:level_crossing}) simplify as
\begin{equation}
      H 
   \approx  \frac{\log 2}{p} + H_c
 \label{eq:level_crossing_approx}
\end{equation}
     As we have mentioned, for a given $H$ the solution of
     Eq.~\ref{eq:level_crossing} determines the $\tilde p(H)$ value of
     the pressure, where the system behaves similarly as presenting a
     first order transition. Hereafter we call $\tilde p(H)$ as the
     level crossing pressure. Alternatively, if we keep the pressure
     fixed and deal with $H$ as a control parameter, from
     Eq.~(\ref{eq:level_crossing}) we get the level crossing value of
     the channel width, $\tilde H(p)$.  From here on we use the
     notations $\tilde \Lambda(p) = \Lambda^+_1(\tilde H(p), p) =
     \Lambda^+_2(\tilde H(p), p)$ and $\tilde V(p) = V^+(\tilde H(p),
     p)$. It is easy to show that $\lim_{H \rightarrow H_c}\tilde
     p(H)= \infty$ and $\lim_{p \rightarrow \infty} (\tilde
     V(p)/\tilde \Lambda(p)) = 0$, which explain why the level
     crossing approximation becomes more valid as $H$ approaches
     $H_c$.

     It is important to note that both the orientational correlation
     length and the compressibility truly diverge as $(p,H)
     \rightarrow (\infty,H_c)$ in such a special way that the
     condition $\Lambda^+_1 = \Lambda^+_2$ is always fulfilled.  When
     we go on this level crossing line, for large $p$ (from which
     follows that $\tilde V \ll \tilde\Lambda$), the orientational
     correlation length goes with $p$ as
\begin{equation}
      \xi_o
   =  \left[ -\log\left(\frac{\lambda_1}{\lambda_0}\right) \right]^{-1} 
   \approx \frac{\tilde \Lambda}{2 \tilde V}
   \approx  \frac{1}{2} e^{p(3/2-\sqrt{2})}, 
 \label{eq:xi_near_critical_point}
\end{equation}
     from where we observe that the orientational correlation length
     diverges exponentially. Further in the text we argue that this is
     really a critical divergence.

\subsection{The scaling property of the orientational correlation function and the Gibbs free energy}

     When we see more carefully the correlation function given by
     Eq.~(\ref{eq:G_o-exp}), we can observe that on the level crossing
     line the amplitude of the correlation function equals one
     ($\Lambda^+_1 = \Lambda^+_2 \Rightarrow S^+ = 0 \Rightarrow 2
     a_1^+ a_2^+ = A_o^{1/2} = 1$). That is, the orientational
     correlation function,
\begin{equation}
      G_o(n) = e^{-n/\xi_o}
 \label{eq:scaling_of_G_in_our_model}
\end{equation}
     depends only on $n/\xi_o$, as it is usual near a
     critical point, where the system shows a scaling behavior. In
     general, the scaling form of the correlation function is written
     as (see e.g.  Ref~\cite{Chaikin-Lubensky} or \cite{Cardy})
\begin{equation}
      G(n)
   =  n^{-(d-2+\eta)} Y(\frac{n}{\xi})
 \label{eq:scaling_of_G_in_general}.
\end{equation}
     By comparing Eqs.~(\ref{eq:scaling_of_G_in_our_model})
       and (\ref{eq:scaling_of_G_in_general}) we must conclude that
     in our case $d-2+\eta = 0$.

     We can now see that in the vicinity of $(p_c, H_c)$ not only the
     orientational correlation function but other physical properties
     show a scaling behavior when expressed in terms of the
     correlation length. This is a common practice in one dimensional
     systems, rather than expressing the properties in terms of the
     reduced temperature or pressure. In the limit of $p \rightarrow
     p_c$ (it follows that $\tilde V \ll \tilde \Lambda$) and $H
     \approx \tilde H \rightarrow H_c$ (it follows that $\Lambda^+_1
     \approx \Lambda^+_2 \approx \tilde \Lambda$ and that
     $|\Lambda^+_1 - \Lambda^+_2|/2 \ll \tilde V$) one can get
\begin{equation}
      \lambda_0
   \approx
      \tilde \Lambda 
      \left(
        1 + \frac{\tilde V}{\tilde \Lambda}
          \left[
            1 + \frac{1}{2}\left( 
                              \frac{\Lambda_1^+ - \Lambda_2^+}{2 \tilde V} 
                           \right)^2
          \right]
      \right)
\end{equation}
and
\begin{equation}
      g
   \approx
     -\log \tilde \Lambda
     -\frac{1}{2} \left( \frac{2 \tilde V}{\tilde \Lambda} \right)
               \left[
                 1 + 2 \left( 
                         \frac{\tilde \Lambda}{2 \tilde V}
                       \right)^2
                       \left( 
                         \frac{\Lambda_1^+ - \Lambda_2^+}{2 \tilde \Lambda} 
                       \right)^2
               \right].
 \label{eq:g_scaling_1}
\end{equation}
     Using Eq.~(\ref{eq:xi_near_critical_point}) this last expression
     can be identified with the general scaling form for the singular
     part of the free energy~\cite{Chaikin-Lubensky,Cardy}
\begin{equation}
      g - g_0
   \sim \xi^{-d} X(h \xi^{\lambda}),
 \label{eq:scaling_of_g_in_general}
\end{equation}
     where $g_0$ is the regular part of $g$, $h$ is some kind of
     external field, and $X$ is a scaling function. By comparing
     Eqs.~(\ref{eq:g_scaling_1}) and
     (\ref{eq:scaling_of_g_in_general}) we find that
\begin{equation}
      g - g_0
   = -\frac{1}{2} \xi_o^{-1} [ 1 + 2 (h \xi_o)^2 ],
 \label{eq:scaling_of_g_in_our_model}
\end{equation}
     where $g_0 = -\log \tilde \Lambda$ and the external field $h$
     measures the distance from the level crossing line defined by
     Eq.~(\ref{eq:level_crossing}),
\begin{align}
      h
  &=  \frac{\Lambda_1^+ - \Lambda_2^+}{2 \tilde \Lambda}
 \label{eq:h_def}
\\
  &\approx
      \frac{1}{2} 
        \left[
          -p(H-H_c) + \log 2 - \log(1+e^{-p(H-\sqrt{2})})
        \right].
 \nonumber
\end{align}

     The scaling property of the correlation function and the free
     energy (Eqs.~(\ref{eq:scaling_of_G_in_our_model}) and
     (\ref{eq:scaling_of_g_in_our_model})), prove that in the vicinity
     of the $(p_c,\ H_c)$ point our model shows critical behavior.
     From Eqs.~(\ref{eq:scaling_of_g_in_general},
     \ref{eq:scaling_of_g_in_our_model}) we must conclude that our
     system really behaves as a one dimensional system, $d=1$;
     moreover $\lambda=1$. The two independent exponents can be chosen
     as $\eta=1$ and $\lambda=1$, and the other usual exponents can be
     calculated from the scaling laws.  Alternatively, we can compute
     directly the scaling of the isothermal compressibility and the
     specific heat.  We observe from
     Eq.~(\ref{eq:scaling_of_g_in_our_model}) that
\begin{align}
      \left.\frac{\partial g}{\partial p}\right|_H(\tilde p)
   =  -\frac{1}{2} X^\prime(0)
       \left.\frac{\partial h}{\partial p}\right|_H(\tilde p)
   =  0,
 \label{eq:g'}
\end{align}
     because $X(a)=1+2a^2$, and therefore $X^\prime(0)=0$. But
     $X^{\prime\prime}(0)=4$, and so the second derivative of $g$ is
\begin{align}
     -\left.\frac{\partial^2 g}{\partial p^2}\right|_H(\tilde p)
   \approx \frac{\xi}{2} (H(\tilde p)-H_c)^2
   \approx \left(\frac{\log 2}{2 \tilde p}\right)^2e^{-\tilde p(\sqrt{2}-3/2)}.
  \label{eq:g''}
\end{align}
     Now, taking into account Eq.~(\ref{eq:kappa}) we can see that
     $\kappa_{_T} \sim \xi$. A comparison with the usual definition of
     the exponent $\bar\gamma = \gamma/\nu$, from which $\kappa_{_T}
     \sim \xi^{\bar\gamma}$, leads to the conclusion that $\bar\gamma
     = 1$.  Similarly, the isobar specific heat (isobar in the sense
     that the longitudinal pressure $p$ is constant, but also $H=const.$)
     can be written as
\begin{equation}
      c_p
   = -T \left.\frac{\partial^2 G}{\partial T^2}\right|_{p_x}
   =  k_B N p \frac{\partial g}{\partial p} 
     -k_B N p^2 \frac{\partial^2 g}{\partial p^2},
\end{equation}
     and therefore, from Eqs.~(\ref{eq:g'}) and ~(\ref{eq:g''}) we
     obtain $c_p \sim \xi$, implying that the exponent $\bar\alpha =
     \alpha/\nu$, such that $c_p \sim \xi^{\bar\alpha}$, equals one.

     By comparing Eqs.~\eqref{eq:level_crossing} and
     \eqref{eq:h_def} one can conclude that $h=0$ means that the
     pressure is $\tilde p(H)$, i.e. the system is at the level
     crossing line where the order parameter is zero. Therefore,
     $\lim_{h \to 0}\langle S_o \rangle = 0$. In other words, there is
     not a spontaneous formation of an ordered phase in our model
     (which is not surprising in quasi-1d).  Consequently, the value
     of $\bar\beta=\beta/\nu$, the critical exponent related to the
     order parameter, cannot be computed directly. However, from the
     scaling laws we can determine its value.  We have $\lambda=1$,
     $\bar\alpha =\bar\gamma = 1$ and $\eta = 1$.  From the scaling
     law $\lambda = \bar\beta + \bar\gamma$ we conclude that
     $\bar\beta=0$.  Alternatively, from
     Eq.~(\ref{eq:scaling_of_g_in_our_model}) we have seen that the
     dimension is $d=1$, and from the scaling law
     $\bar\beta=\frac{1}{2}(d-2+\eta)$ we consistently obtain
     $\bar\beta=0$. The effect of $\bar\beta=0$ is observed in
     Fig.~\ref{fig:<S_o>--p}, that is, as $\tilde p \rightarrow p_c$
     on the level crossing line, both $\langle S_o(H)
     \rangle|_{p=const.}$ and $\langle S_o(p) \rangle|_{H=const.}$ go
     to the step function. This is a very important difference
     compared with the usual critical points in three dimensions.

     From the obtained critical exponents we can see that \emph{the
       $(p_c, H_c)$ critical point belongs to the universality class
       of the 1d Ising model.} The important consequence of this
     result, as we discussed in the previous paragraph, is that the
     critical exponent $\bar\beta=0$ (contrasting with the usual
     $\beta=1/3$ value of 3d systems), implies that the order
     parameter is discontinuous at the $(p_c, H_c)$ point.  Despite
     the fact that this discontinuity disappears at any finite
     pressure, it has a significant impact on the system behavior near
     the critical point.  Namely, the order parameter and the density
     behave almost like the step function, and their derivatives, as
     the compressibility, have high peaks.  Simulation results, as
     close as possible to the $(p_c, H_c)$ point and for the
     unrestricted $y$ and $\varphi$ system, show that the system
     behavior is indistinguishable from that of a first order
     transition~\cite{Gurin_PRE_2016}.
     
     The discontinuity can disappear at any finite $p$ because the
     thermodynamic quantities are singular at the $(p_c, H_c)$ fixed
     point, contrary to the case of a zero temperature discontinuity
     fixed point of the usual 3d systems, like the ferromagnets. In
     this last case the thermodynamic quantities are not singular at a
     discontinuity fixed point which separates the different phases,
     and perturbation theory should have a finite radius of
     convergence, therefore the reason of the discontinuity, the
     coexistence of the different phases must persist for some
     distance into the phase diagram. This coexistence line is
     terminated at a different, critical fixed point~\cite{Cardy}. In
     1d, these two fixed points, the critical and the discontinuity
     ones are merged in a unique fixed point. This is the origin of
     the peculiar behavior of our 1d system, which is reminiscent of a
     first order phase transition though every property can be
     expressed as an analytic function.

     Let us now discuss the physical origin of the ``external field''
     given by Eq.~(\ref{eq:h_def}). By extending our model with an
     external field which favors the $\Diamond$ orientation against
     with the $\square$ one, including in the Hamiltonian a $-
     h^\prime \sum_i \cos(4\varphi_i)$ term, this extra field
     $h^\prime$ simply appears as an additive term in
     Eq.~(\ref{eq:h_def}). This points out that $h$ has exactly the
     same effect as an extra external field $h^\prime$ which is
     directly coupled to particle orientation. But in our system the
     special combination of two different external fields---the
     external pressure $p$ and the wall of the channel---results in an
     effective field which favors one orientation or the other.
     Moreover, the strength and the ``direction'' of this effective
     field depend on the values of $p$ and $H$, given that the
     longitudinal pressure favors the closely packed $\Diamond$
     orientation (having vanishing $y$-fluctuations) and the wall the
     $\square$ orientation. In other words, the direction of this
     external field results from an entropic competition between the
     $x$ and $y$ fluctuations.
    
     We would like to emphasize the following interesting feature of
     the $(p_c,\ H_c)$ point: the compressibility diverges as $p
     \rightarrow \infty$. This is quite special in a system consisting
     of only rigid particles. Usually $p \rightarrow \infty$ implies
     that the system approaches the close packing structure while the
     compressibility goes to zero.  But in our system, at $(p_c,\
     H_c)$ the close packing structure is degenerated, because the
     $\square$ and the $\Diamond$ orientations have the same close
     packing density. This is the key feature of this point: the
     system cannot decide between these two competitive structures. In
     this sense, it is very similar to a spontaneous symmetry
     breaking. Note that other systems may show this peculiar point.
     For instance, hard anisotropic particles get spatially and
     orientationaly ordered at close packing, whereas spheres (or
     disks) get only spatially ordered. In the limit of small
     anisotropy, both, the orientationaly ordered and disordered
     structures also have the same packing fraction, and the structure
     gets again degenerated ~\cite{Bautista13,Bautista14}.

\subsection{The $y$ positional correlations}

     Up to this point we have focused only on the orientational
     correlations in our model, which can be totally described by the
     $2 \times 2$ transfer matrix $K_+$ given by
     Eq.~(\ref{eq:blocks_K}). Thus, the analogy with the 1d Ising
     model comes as no surprise because this last model can be
     described by a $2 \times 2$ transfer matrix, too.  However, in
     addition to the continuous longitudinal translational degrees of
     freedom, every particle has 4 discrete possible states, instead
     of the 2 states of the Ising model. 

\begin{figure}[h!]
   \subfigure[]
   {
      \includegraphics{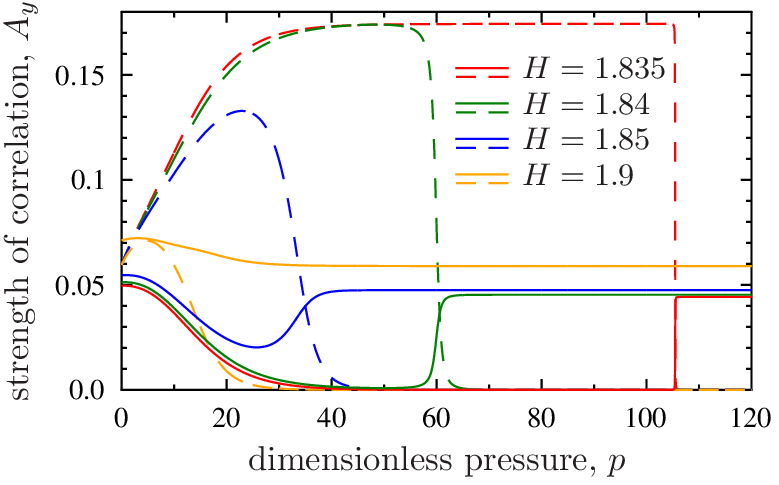}
      \label{fig:A_y--p}
   }
   \subfigure[]
   { 
      \includegraphics{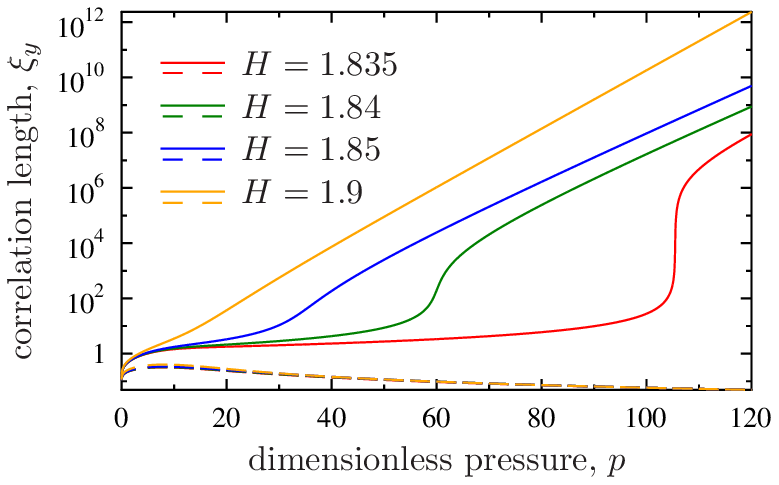}
      \label{fig:xi_y--p}
   }
   \begin{center}
   \end{center}
   \caption{The positional correlation function, see
     Eq.~(\ref{eq:G_y}) and Eq.~(\ref{eq:G_y-exp}). The solid lines
     represent $A_y$ and $\xi_y$, the dashed lines are $A_y^\prime$
     and $\xi_y^\prime$, the later ones are very small and practically
     coincide for all channel widths.
\label{fig:G_y--p}}
\end{figure}

     To evaluate the $y$ positional correlation function we have to
     take into account the 2nd and 3rd eigenvalues and eigenvectors of
     the transfer operator, and so the above mentioned $2 \times 2$
     matrix, $K_+$, is not enough anymore. The expectation value $\langle 
y\rangle$ is zero, as we have mentioned at the end of Sec.~\ref{sec:II}.
     Its correlation function, according to Eq.~(\ref{eq:G_y}), can be
     written as
\begin{equation}
      G_y(n)
   =  (-1)^n A_y e^{-n/\xi_y} + A_y^\prime e^{-n/\xi_y^\prime}, 
 \label{eq:G_y-exp}
\end{equation}
     and the corresponding results are depicted in
     Fig.~\ref{fig:G_y--p}.  This last equation is again exactly valid
     for any neighboring distance $n$. The amplitude of the $y$
     correlations, $A_y = (a_1^+ a_2^- \frac{H-1}{2} - a_2^+ a_1^-
     \frac{H-\sqrt{2}}{2})^2$ and $A_y^\prime = (a_1^+ a_1^-
     \frac{H-1}{2} + a_2^+ a_2^- \frac{H-\sqrt{2}}{2})^2 $ are shown
     in Fig.~\ref{fig:A_y--p}. The correlation length, $\xi_y =
     (-\log(|\lambda_3|/\lambda_0))^{-1}$ and also $\xi_y^\prime =
     (-\log(\lambda_2/\lambda_0))^{-1}$ are shown in
     Fig.~\ref{fig:xi_y--p}.

     From Fig.~\ref{fig:A_y--p} it can be seen that the positional
     correlations show mainly a non-alternating behavior when
     $p<\tilde p(H)$, but suddenly become alternating for
     $p > \tilde p(H)$. The picture is as follows: when the
     orientation of the particles is $\square$, then the correlation
     is weak, non-alternating, and short range (nevertheless, the
     alternating part has relatively long correlation length and small
     amplitude, which is due to the presence of few $\Diamond$
     particles.)  When the orientation of the particles is mainly
     $\Diamond$, then the correlation is strong, alternating, and long
     range.

     It is interesting that the correlation length $\xi_y$ suddenly
     increases at $\tilde p$ but has no peak; it is a monotonic
     function and diverges exponentially with $\tilde p$. This kind of
     divergence has been also observed in other systems of hard body
     particles such as rectangular~\cite{Kantor_EPL_2009} or
     V-shaped~\cite{Khandkar_PRE_2017} particles confined to a line.
     The divergence of $\xi_y$ can be explained easily as follows.
     If---due to a fluctuation---a particle changes its position in
     the $\Diamond$ oriented zig-zag phase, then the neighboring
     particles are forced to follow that change to avoid forming
     domain walls in the zig-zag structure.  This means that the
     correlation among the $y$ positions is strong.  This effect
     becomes stronger with increasing pressure because the cost of
     domain walls increases, therefore the correlation length
     increases, too.

     It is interesting
     to see that at the level of the transfer operator the reason of
     this sudden change of $\xi_y$ is due to the fact that $\lambda_0
     > |\lambda_3| > \lambda_1$, where $\lambda_0$ and $\lambda_1$
     produce an avoided level crossing, thus $|\lambda_3|$ gets stuck
     between $\lambda_0$ and $\lambda_1$. The orientational and $y$
     positional degrees of freedom are not ``entangled'', but the
     orientational level crossing has some impact on the positional
     behavior.
   
     Finally, in our opinion, this divergence of $\xi_y$ is not so
     interesting as the divergence of $\xi_o$, because the latter
     causes an interesting behavior at finite pressure (the peculiar
     behavior of the equations of state, see Fig.~\ref{fig:eta--p})
     and also at infinite pressure (diverging compressibility in spite
     of dealing with a hard system), but the former has not such
     consequences. The reason for this is that the orientational order
     is coupled to both physical external fields, the wall and the
     longitudinal pressure---see Eq.~\ref{eq:h_def}---, while the
     alternating $y$ positional order is not. The zig-zag positional
     order would be coupled to an alternating external field, i.e. the
     Hamiltonian should contain a $h^{\prime\prime} \sum_i (-1)^i y_i$
     term, which is physically unusual, $h^{\prime\prime}$ cannot be
     tuned by $p$ or $H$, therefore the related quantities
     (susceptibility etc.)  are uninteresting.

     \section{Summary and further discussions}

     We have shown that near the $H_c=2\sqrt{2}-1$, $p_c=\infty$ point
     the Gibbs free energy and the orientational correlation function
     show a scaling behavior, so this is a critical point. We have
     calculated all the critical exponents and we have found that our
     model is in the same universality class as the 1d Ising model.
     One can argue that there is no real critical point at any finite
     pressure, however, there is a real (experimentally observable)
     critical behavior in the vicinity of the critical point, which is
     indeed located in the physically meaningless (experimentally
     unreachable) parameter regime ($p_c=\infty$). At this fixed point
     the critical exponent $\bar\beta=0$, implying that the order
     parameter has a discontinuity. This means that this fixed point
     unifies the feature of a usual critical fixed point and a
     discontinuity fixed point of three dimensional systems.
     Therefore, in the vicinity of this point the system behaves very
     similar as showing a first order transition, however, at the same
     time, the peaks of the compressibility and the specific heat are
     typically like that of a critical system. The free energy is
     singular at $(p_c, H_c)$, therefore the discontinuity disappears
     at any finite $p$, turning all thermodynamic functions analytic.

     In spite of the similarities between our system and the 1d Ising
     model, we want to emphasize some differences in the underlying
     physics. We have mentioned that in our model every particle has
     continuum translational $x$ degrees of freedom and four possible
     discrete states while the Ising spins are localized and have only
     two different discrete states, but more important, in the Ising
     model the thermal fluctuations can change the direction of a
     \emph{single spin alone} at any temperature. Certainly, when the
     temperature is small, the probability of spin flipping is very
     small too, but possible, irrespectively to the states of its
     neighbors. In our model the "flipping" of a square from the
     rotational state parallel to the wall ($\square$) into the other
     state when its diagonal is parallel to the wall ($\Diamond$) is
     impossible at high densities without the disturbance of its
     neighbors. At high densities fluctuations can "flip" a square
     only together with many other neighbors, what is more, the
     positions of the squares have to be changed at the same time.
     \emph{Only collective motions can change the orientational state
       of the particles}, which makes an important difference.

     We would like to add that this kind of model, namely hard
     particles confined into a narrow, quasi-one-dimensional channel,
     is often regarded as a simple model to study the glassy or
     jamming phenomenon, see for example Refs.~\cite{Bowles_PRL_2012,
       Bowles_PRL_2013, Godfrey-Moore_PRE_2015,
       Godfrey-Moore_PRE_2016}.  Most of these works focus on disks
     confined by somewhat wider channels than ours, but the observed
     glassy behavior is similar since only collective rearrangements
     of particles can increase the density, and with increasing
     pressure these collective rearrangements become less probable;
     the system is sticked into the so called locally jammed states.
     These works report interesting features such as an isobaric heat
     capacity maximum and a corresponding diverging pressure for some
     density below the maximal packing fraction.  Then, they conclude
     that these features are related to some kind of fragile-strong
     fluid crossover or avoided phase transition, which is
     phenomenologically reminiscent of the bulk glass transition. Here
     we would like to provide an alternative point of view. We think
     that the almost singular behavior of the thermodynamic quantities
     \emph{below} the close packing density can be a marker of a
     critical behavior related to a critical point \emph{at} the close
     packing density. The system studied in our work gives an example
     for this possibility. The existence of a critical point at
     $(p_c,H_c)$ has consequences at finite $\tilde p$ pressures for
     $H>H_c$, namely: $c_p$ and $\kappa_T$ have large peaks, and the
     pressure goes up suddenly at a given (below the close packing)
     density. These features are very similar to those reported in
     connection with the fragile-strong fluid crossover or avoided
     phase transition. This behavior can be explained coherently by
     the existence of a fixed point at infinite pressure, which
     unifies the properties of a usual critical fixed point and a
     discontinuity fixed point.

     It remains an open question what is the relationship between the
     present criticality and the possible glass/jam behavior. Here we
     would like to point out the differences. First of all, it is a
     long standing question whether the jamming transition has some
     sign in the equilibrium properties of the system or not. We
     emphasize that we studied only equilibrium properties. On the
     other hand, to understand the critical behavior in our model it
     is enough to take into account only two competitive structures.
     This contrasts with a glass, where the system have not only two,
     but many (usually very much) almost stable but actually
     metastable states.  Moreover, in our system, the reason of the
     existence of metastable states is the presence of the confining
     walls, which strongly decrease the room for rearranging
     configurations.  Finally, the existence of a merged critical
     \emph{and} discontinuity fixed point is typical in one dimension.
     Therefore, the extrapolation of the conclusions deduced from such
     quasi-1d systems to bulk 3d systems is, from our point of view,
     strongly questionable.

     Another issue is the analogy between the present and the
     continuous models. How relevant are the presented results for the
     freely rotating and moving case, when all the degrees of freedom
     are continuous? We have no exact answer, but we strongly believe
     that for the case where the $y$ degrees of freedom are continuous
     with discretized orientation, the long range orientational
     properties of our model can be effectively described by a $2
     \times 2$ matrix and, as a consequence, the 1d Ising like
     critical point is preserved.  But even in this case the system
     cannot be completely resolved so easily.  As we have mentioned,
     the orientational and $y$ positional degrees of freedom turn
     ``entangled'' and it becomes not so trivial to construct the
     effective $2 \times 2$ matrix. On the other hand, treating the
     continuous rotational degrees of freedom is mathematically more
     subtle.  Nevertheless, physical considerations suggest that the
     underlying reason for the ``almost'' singular behavior at
     pressure $\tilde p(H)$ is not the discrete nature of the ($y$ and
     orientational) degrees of freedom. What it is more, the extra
     degrees of freedom enhance the singular-like behavior even for
     not so strong confinement.  In the discrete system studied in
     this paper there is nothing interesting when $H>1.9$, and the
     curves just start to show singular-like behavior below $H=1.85$,
     see e.g.  Fig.~\ref{fig:eta--p} or \ref{fig:compressibility--p}.
     We know that the situation is more exciting in wider channels
     when $y$ is continuous~\cite{Gurin_PRE_2016}. The reason is that
     the negative constant part of the external field $h$ in
     Eq.~(\ref{eq:h_def}) comes from the $y$ positional fluctuations.
     When $y$ is continuous, fluctuations have more room as the phase
     space is larger. Therefore $h$ is more negative and favors more
     strongly the $\square$ orientation.  The emerging picture is as
     follows: $y$ fluctuations favor the $\square$ orientation.
     Therefore, by increasing the number of $y$-degrees of freedom the
     ``transition'' pressure, $\tilde p$, goes up, approaching the
     critical point. Our numerical results show~\cite{Gurin_PRE_2016}
     that the freely rotating case shifts its ``transition'' pressure
     further more to high values, although in this case it is not
     trivial the reason why. In this unrestricted system we have no
     prove of the existence of a critical point, but we strongly
     believe on its presence.


%

\end{document}